\documentclass[acmsmall, nonacm]{acmart}
\AtBeginDocument{%
  }

\setcopyright{cc}
\setcctype[4.0]{by}
\acmYear{2024} 
\copyrightyear{2024}
\acmVolume{1} 
\acmNumber{FSE} 
\acmArticle{32} 
\acmMonth{7}
\acmDOI{10.1145/3643758}

\begin{document}

\title[Rocks Coding, Not Development]{Rocks Coding, Not Development--A Human-Centric, Experimental Evaluation of LLM-Supported SE Tasks}

\title{Rocks Coding, Not Development: A Human-Centric, Experimental Evaluation of LLM-Supported SE Tasks}

\author{Wei Wang}
\orcid{0000-0003-3240-343X}
\affiliation{%
  \institution{Beijing University of Posts and Telecommunications}
  \city{}
  \country{China}
}
\email{weiwang@bupt.edu.cn}

\author{Huilong Ning}
\orcid{0009-0001-9393-6507}
\affiliation{%
  \institution{Beijing University of Posts and Telecommunications}
  \city{}
  \country{China}
}
\email{nulogn@bupt.edu.cn}

\author{Gaowei Zhang}
\orcid{0009-0006-5767-2280}
\affiliation{%
  \institution{Beijing University of Posts and Telecommunications}
  \city{}
  \country{China}
}
\email{zhanggaowei@bupt.edu.cn}

\author{Libo Liu}
\orcid{0000-0002-0136-8902}
\affiliation{%
  \institution{University of Melbourne}
  \city{}
  \country{Australia}
}
\email{libo.liu@unimelb.edu.au}

\author{Yi Wang}
\orcid{0000-0003-1321-4035}
\affiliation{%
  \institution{Beijing University of Posts and Telecommunications}
  \city{}
  \country{China}
}
\email{wang@cocolabs.org}

\renewcommand{\shortauthors}{W. Wang, H. Ning, G. Zhang, L. Liu, Y. Wang}

\begin{abstract}
Recently, large language models (LLM) based generative AI has been gaining momentum for their impressive high-quality performances in multiple domains, particularly after the release of the ChatGPT. Many believe that they have the potential to perform general-purpose problem-solving in software development and replace human software developers. Nevertheless, there are in a lack of serious investigation into the capability of these LLM techniques in fulfilling software development tasks. In a controlled 2 $\times$ 2 between-subject experiment with 109 participants, we examined whether and to what degree working with ChatGPT was helpful in the coding task and typical software development task and how people work with ChatGPT. We found that while ChatGPT performed well in solving simple coding problems, its performance in supporting typical software development tasks was not that good. We also observed the interactions between participants and ChatGPT and found the relations between the interactions and the outcomes. Our study thus provides first-hand insights into using ChatGPT to fulfill software engineering tasks with real-world developers and motivates the need for novel interaction mechanisms that help developers effectively work with large language models to achieve desired outcomes.  
\end{abstract}

\ccsdesc[500]{Human-centered computing~Laboratory experiments}
\ccsdesc[500]{Software and its engineering~Software development techniques}
\ccsdesc[500]{Software and its engineering~Software development methods}

\keywords{large langauge models, controlled experiment, software development task, human-AI collaboration}

\maketitle

\section{Introduction}
Artificial Intelligence Generated Content (AIGC) technologies, particularly the large language models (LLM), have been gaining momentum since the release of ChatGPT\footnote{ChatGPT: \url{https://chat.openai.com}.} in December 2022. Given their remarkable performances in generating human-level content, such LLMs and related tools were widely assumed to have the capability to be game changers for many knowledge professions \cite{van2023chatgpt}, including software development. There were some claims that professional software development may soon be replaced by large language models that can generate code automatically according to the prompts. For example, Matt Welsh, CEO of Fixie.ai, delivered a talk titled ``\emph{Large Language Models and The End of Programming}'' on May 9th \cite{10.1145/3570220}. Such claims were often backed by ChatGPT's performances in solving algorithm/coding puzzles in online judging platforms such as \textsc{LeetCode} or passing algorithm/coding interviews of big techs.  

However, software development is much more than solving algorithm/coding puzzles. It involves various activities in its life-cycle, to name a few, requirements elicitation, design, implementation, maintenance, project management, etc. \cite{10.1145/3387111}. Even the narrowly-defined development--code production--contains much more than writing a small program to solve coding puzzles. Take the \textsc{GitHub} commit as an example; there were six types of activities at the source code file level \cite{9726925}, among which \emph{MODIFY} rather than others accounts for 81\%, an overwhelming majority among all activities. Thus, real-world software development task often requires reading, understanding, and modifying the existing code. Even supposing professional software engineers were willing to use LLM-based generative AI techniques, it is unclear whether using such tools would benefit their productivity and how their work practices would be affected when they need to collaborate with such tools. For example, when discussing natural language to code tools, Xu et al. \cite{10.1145/3487569} argued that ``\emph{there is a surprising paucity of research using \textbf{human-centered approaches}} \cite{myers2016programmers} \emph{to evaluate the usefulness and impact of these methods within the software development workflow}.'' Besides, the AI community usually relies on automated evaluations on benchmarks without considering human involvement. Thus, we must be cautious about the high performances of AI models when applying them in real-world software development.   

We sought to fill such a gap by empirically examining the effects of involving ChatGPT in performing typical software engineering tasks. The first natural concern was whether working with ChatGPT would be helpful in performing different tasks, and, how helpful it would be, in terms of the efficiency of performing such tasks, their solutions' quality, and subjective task load. Thus, we have the first research question. 

\begin{description}
\item[$\mathbf{RQ_1}$:] \emph{Does working with ChatGPT bring benefits to the outcomes in performing coding and software development tasks? If yes, to what extent could be such benefits of working with ChatGPT?}
\end{description}
Moreover, introducing ChatGPT--a novel and unfamiliar LLM tool into developers' software development workflow, would unavoidably change those developers' behaviors. To some degree, developers have to find a way to collaborate with these tools. Thus, our second research question is formulated as follows: 
\begin{description}
\item[$\mathbf{RQ_2}$:] \emph{How do people interact with ChatGPT during their collaboration in fulfilling tasks?}
\end{description}

With 109 participants, we conducted a 2 $\times$ 2 between-subject controlled experiment to answer the above research questions. By assigning participants into different conditions, we could compare the task outcomes with(out) ChatGPT's support in different tasks. We also observed how participants interacted with ChatGPT in fulfilling the tasks. Our findings revealed that while ChatGPT was exhibit certain superiority in helping developers solve coding puzzles, its capabilities in supporting typical development task were limited, no matter in work efficiency and quality of solutions. Furthermore, we characterized the developers' interaction with ChatGPT in their collaboration, and found rich dynamics. Our findings rebutted the claim that LLMs would soon replace software developers and emphasize the collaboration between human and LLMs. Based on the findings, we discussed the study's implications for SE research, practices, and education. Thus, the contribution of this paper is tri-fold. 
\begin{enumerate}
    \item We provided a quantitative evaluation of the capability of LLMs, represented by the ChatGPT, in solving coding and typical software engineering tasks. As far as our best current knowledge, it is the first work employing large-scale controlled experiment for evaluating LLMs capabilities in the SE domain. 
    \item We summarized a qualitative description of how people interact with LLMs, represented by the ChatGPT, in solving coding and typical software engineering tasks. 
    \item We identified and discussed a set of implications for future software engineering research, practices, and education when working with LLMs.
\end{enumerate}

\noindent\emph{Paper Organization.} The rest of this paper proceeds as follows. \S2 briefly introduces the background and related work. \S3 presents the experiment design. \S4 reports the results of the experiment. \S5 discusses the implications to software engineering research, practices, and education. \S6\&\S7 discusses the threats to the validity and concludes the paper, respectively.

\section{Background and Related Work}

\subsection{AIGC and LLM}
AIGC models were not new. Early models such as Hidden Markov Models could be dated back to the early 1950s. However, these models could only generate sequential data but not others. Indeed, deep learning gave the second birth of generative AI in both natural language processing (NLP) and computer vision (CV) areas. In NLP, models such as RNNs \cite{mikolov2010recurrent}, LSTM \cite{hochreiter1997long}, and GRUs \cite{chung2014empirical} had been applied in text generation tasks. Deep learning models could generate longer content than simple N-gram statistical learning models. In CV, the proposal of GAN marked substantial progress in image generation tasks \cite{goodfellow2020generative}. VAEs and similar diffusion generative models had also been developed for more fine-grained control over the image generation process. They usually achieved the ability to generate high-quality images. Stable diffusion \cite{borji2023generated}, an instance of latent diffusion models, becomes very popular lately.

The transformer architecture received both communities' interests. A research team from Google invented a transformer to deal with NLP tasks in 2017 \cite{vaswani2017attention}. Transformer architecture has become the dominant choice in NLP due to its parallelism and learning capabilities, facilitating the flourishing of large language models. Many such LLMs, e.g., BERT, GPT, and ChatGPT, used the transformer architecture as their primary building block \cite{rothman2022transformers}. Recently, joint vision-language models such as CLIP \cite{radford2021learning} were proposed to combine the transformer architecture with visual components. Thus, it was trained on a huge amount of text and image data. Since it combined both visual and language knowledge during the pre-training phase, it could also be used as an image encoder in prompting for image and text generation, which realized the multimodality. To sum up, the emergence of transformer-based models revolutionized AIGC and led to the possibility of large-scale pre-training.

Recently, LLMs were increasingly combined with RLHF (Reinforcement Learning from Human Feedback) \cite{ouyang2022training}. RLHF was a technique to achieve the alignment between human preferences and AI-generated content. A pre-trained model on large-scale datasets was built first, then, a reward model was trained to encapsulate the human preferences. By fine-tuning the initial language model to maximize the learned reward function using reinforcement learning, both performances and human preferences were guaranteed. ChatGPT and GPT-4 were exactly built in this way.

\subsection{Human-AI Collaboration}
Our work also broadly connects with the emerging topic of human-AI collaboration which originated from the concept of symbiotic computing~\cite{symbiosis} in 1960. Human-AI collaboration referred that human and AI systems have mutual goal understanding, co-manage the task and share the progress tracking. With this term, incorporating human-centered design philosophy is crucial for successfully designing human-AI working systems that can collaborate with us instead of informing us by outputing something from ``smart black boxes". Human-centered design puts the focus on understanding the needs, motivations, emotions, behavior, and perspective of the people in the development of a design, rather than solely on technical capabilities~\cite{HCA}. 

Since 1960, human-AI collaboration systems has experienced persistent growth, contributing innovative solutions to a diverse array of application fields, such as AI-powered clinical decision support systems in medicine~\cite{medicine}, customer service chatbots in customer service~\cite{service}, and automated machine learning (AutoAI) in data science~\cite{wang2019human}. Recently, LLMs had been experiencing fast growth, they may exhibit issues such as biases, opacity, and insufficient controllability in real-world applications. To address these challenges, researchers are focusing on employing human-AI collaboration to enhance the effectiveness of LLMs  in complex tasks. For instance, by improving auditing tools~\cite{auditing}, introducing chained operations and incorporating modular designs~\cite{chains}, and presenting new collaborative datasets~\cite{dataset}, these studies provide guidance for developing safer, fairer, and more controllable Human-AI collaboration systems. Focusing on AI-assisted coding, some empirical studies have revealed some phenomena in human-AI collaboration. Do users write more insecure code with AI assistants? No, users who had access to an AI assistant based on codex model wrote significantly less secure code than those without access \cite{perry2022users}, the use of LLM does not introduce new security risks. \cite{sandoval2023lost}. Copilot, an AI-based code assistant, did not necessarily improve the task completion time or success rate, but users preferred to use it, since Copilot often provided a useful starting point and saved the effort of searching online \cite{vaithilingam2022expectation}. Also, the rate with which shown suggestions are accepted, drives developers’ perception of productivity \cite{ziegler2022productivity}.

\subsection{LLM and Software Engineering}
Long before the emergence of LLM, using AI technologies to generate software source code had existed. Early studies such as Price et al. \cite{price} had attempt to translate natural language to source code, through with very limited scope. Specifically, they proposed NaturalJava, a prototype of an intelligent user interface for creating, modifying, and examining Java programs. Subsequently, Vadas et al.~\cite{vadas} introduced a system that took unrestricted English as input and gave output code in Python. Later, example-based interactive models like WREX~\cite{wrex} were proposed to generate easily readable code, enabling more efficient data wrangling processes. To further address the mapping problem between natural language descriptions and source codes, some researchers explored neural architectures~\cite{NeuralModel2,NeuralModel1} based on grammar models to capture the syntactic structure of target programming languages, thereby generating higher-quality codes.

With the continuous development of LLMs, they had begun to be applied in more general code generation tasks and brought certain progress. Specifically, inheriting from BERT, CodeBERT~\cite{Codebert} and GraphCodeBERT~\cite{graphcodebert} have powerful code understanding capabilities, achieving state-of-the-art performance in multiple code-related tasks, such as code documentation generation, code translation, code refinement and so on. Codex~\cite{Codex} is a GPT model fine-tuned on publicly available code from GitHub and demonstrates its capabilities in writing Python code. ATHENATEST~\cite{ATHENATEST} and PLBART~\cite{PLBART} are two approaches based on BART~\cite{bart} which combined the characteristics of the bidirectional encoder from BERT and the left-to-right decoder from GPT. ATHENATEST focuses on generating unit test cases and PLBART provides effective solutions on a variety of programming language understanding and generation tasks. Tools for leveraging LLMs in end-user development were also development \cite{10.1145/3544548.3580817}. Meanwhile, ethnographic and experimental studies such as Weisz et al. \cite{10.1145/3397481.3450656} also demonstrated that the generated code was not necessarily to be perfect. Developers could accept imperfect code and work to improve it. Meanwhile, Liang et al.'s recent large scale survey showed that cognitive load of using such tools is still high, i.e., developers often have trouble controlling the tool to generate the desired output addressing certain functional or non-functional requirements \cite{liang2023understanding}. 

\section{Experiment Design}

We followed the standard guidelines of experiment design \cite{jackson2013principles, kirk2012experimental,winer1971statistical} in social and behavioral sciences, and combined the best practices in software engineering \cite{basili1986experimentation,jedlitschka2008reporting, juristo2013basics, 10.1109/TSE.2005.97,wohlin2012experimentation} to design the experiment.

\subsection{Experiment Tasks and Group Setup}
\begin{figure*}
    \centering
    \includegraphics[width=\textwidth]{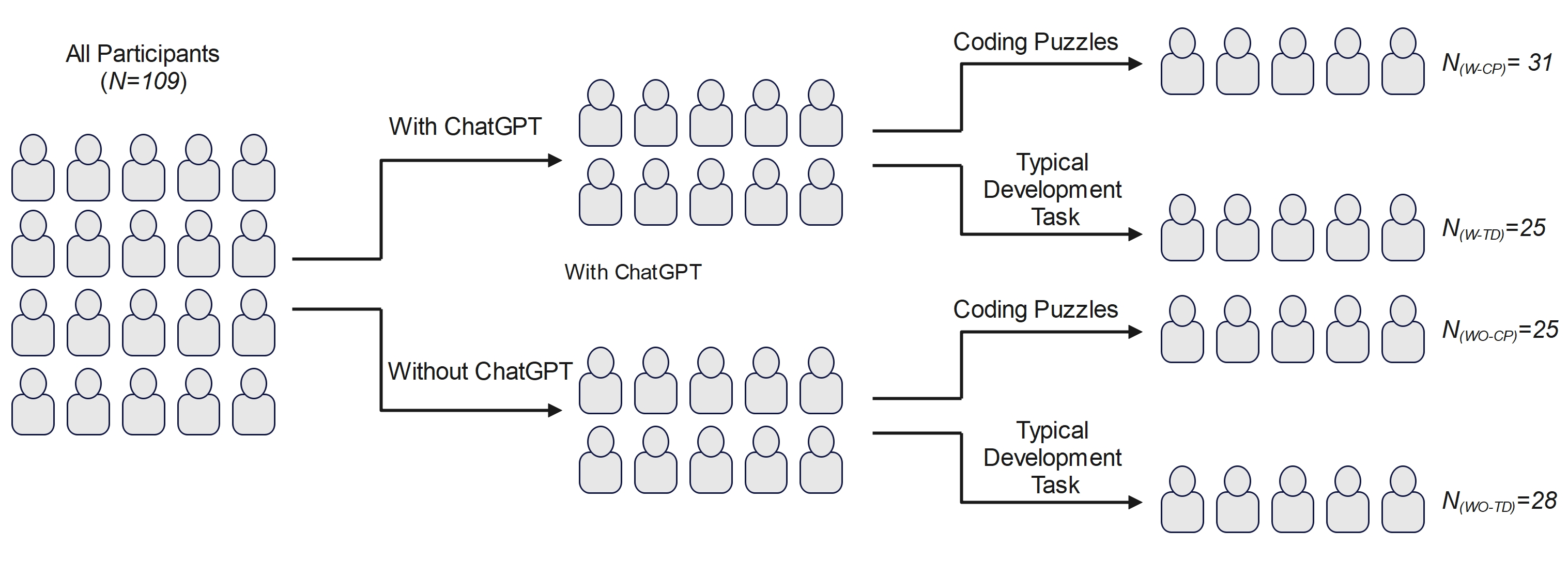}
    \caption{The overall view of the experiment conditions and groups.}
    \label{exp-de}
    \vspace{-1em}
\end{figure*}
\subsubsection{Experiment Tasks}

There were two tasks in the experiment. The first task (task 1) was solving simple coding puzzles. The coding questions on the PTA platform had been classified into five levels from 1 to 5 according to their difficulties from the easiest to the most difficult questions. These questions' difficulties were initially provided by the authors and then automatically adjusted according to the performances of the users by the platform considering many factors such as, ratio of users who solve the problem, average no. of attempts, and time spending on a problem, etc. We created a pool of five coding puzzles (see APPENDIX A) at the same difficulty level of 3. Two questions would be randomly drawn from the pool when a participant was assigned to do this task. Using two randomly-chosen questions rather than one question reduced the chances that a participant may accidentally practiced some coding puzzles, and also made the whole efforts comparable to the second task. The participant would have up to 75 minutes to finish the two coding questions. During this process, the participant could use resources on the Internet (e.g., algorithms, code snippets, etc.) for references. We did not prohibit ``copy and paste'' practices since they had been common parts of daily software development. Participants were asked to finish it in an online programming platform that provided automated judges to the programs written by our participants.  

The second task was used to simulate a real-world small-scale software development task. In this task, a participant was asked to fix two bugs in a small project selected from \textsc{GitHub}. Though real world software development are not merely fixing bugs, bug-fixing task represented a large proportion of software development activities. For example, Yue et al. found that the ``MODIFY'' activities account for over 80\% source code contributions on GitHub \cite{9726925}. Moreover, bug-fixing requires multiple types software development capabilities, for example, coding, testing, programming comprehension, domain knowledge, and so on. It reflects one's all-round abilities in software development. The selected project was an application\footnote{\url{https://github.com/yangxuanxc/wechat_friends}} that crawled a number of WeChat users' profile pictures with their usernames and visualized these pictures properly. The bugs to be fixed were reported as real issues. Fixing such issues requires modifying a few lines of code and testing if the issue was resolved. The fixes for the two bugs were all local, and the participant did not need to change more than one file for each bug. We established a local development environment for participants to finish this task. The source code used in the experiment is made available in the replication package. 

These two tasks were not arbitrarily decided. We chose them based on several rationales: (1) the tasks had to be within a specific scale allowing participants to finish in a reasonable time slot. (2) the first task should at least require a medium level of problem-solving ability; and (3) the second task must be real-world issues. In the rest of the paper, we simply referred task 1 as ``\emph{Coding Puzzles}'' and task 2 as ``\emph{Typical Development Task}.'' Note that although we did not specify which language to use in the study, all participants chose \textsc{Python} in \emph{Coding Puzzles}. Since the source code of the \emph{Typical Development Task} was written in \textsc{Python}, all participants had to use it when fixing the bugs.

\subsubsection{Experiment Group Setup}
Each task was performed under two different conditions, \emph{with} and \emph{without} the assistance of ChatGPT in performing the task. We used a between-subject design, i.e., a participant only needed to finish one task. The experiment thus took the form of the 2 $\times$ 2 factorial design, as shown in Fig. \ref{exp-de}. Each group corresponded to one of the four combinations: (1) With ChatGPT and \emph{Coding Puzzles}, represented by \emph{W-CP}; (2) With ChatGPT and \emph{Typical Development Task}, represented by \emph{W-TD}; (3) Without ChatGPT and \emph{Coding Puzzles}, represented by \emph{WO-CP}; and (4) Without ChatGPT and \emph{Typical Development Task}, represented by \emph{WO-TD}.

Participants were randomly assigned to these four groups. Their group assignments were determined by a random number generated according to their arrival times. No human intervention was involved in the group assignment, thus satisfying the experiment's requirement on the completely randomized design \cite{rubin1980randomization}. Note that the random number generator was not perfect enough to place participants evenly into four groups, so there were slight variances among different groups (See the right part of Fig. \ref{exp-de}). 

\subsection{Experiment Environment}
We implemented a simple web-based platform to manage the study process and collect some information, e.g., a few demographics and post-experiment questionnaires. For each task, we established different experiment environments. Task 1 (Coding Puzzles) participants were guided to an online judge platform (PTA\footnote{PTA:\url{https://www.pintia.cn}}) to finish the task. If they were in the group with ChatGPT, they would be given accounts to use it. Otherwise, they would finish the task directly. A local \emph{dev} environment for Task 2 (typical development task) was built and deployed virtually, allowing participants to edit and test their code so that participants would not need to worry about setting up environments. Of course, accounts of ChatGPT with GPT-3.5 were also ready for use for those in that group. 
To simulate the everyday development practices, Internet access remained available all the time during the experiment for all participants. I.e., using internet resources to finish the tasks, even directly copying from there, was not prohibited because it was a natural part of modern programming \cite{liu2021opportunities}.

\subsection{Participants}

The participants were recruited from a mid-size public university focusing on Information and Communication Technology in Beijing, China. In the remainder of the paper, we were going to use UICTBC to refer to it. The university had one of the largest student populations ($\approx$10K) majoring in computing programs, e.g., computer science, software engineering, artificial intelligence, data science, etc., around the world. The university was viewed as an elite ICT university in China. Most of its graduates joined the technology sector as engineers in the areas of communication technology, electronics, hardware systems, and software development.  

The study was advertised on campus in multiple ways. In total, there were 323 filling in the recruitment form. We selected those majoring in computing areas and had at least one year professional software development experience (mostly as interns in professional software development organizations) to participate in this study. A WeChat group was established to communicate with the participants to schedule the experiment sessions. In total, there were 109 participants. Among them, 89 were male, and 20 were female; 7 were in undergraduate programs, while 102 were in postgraduate programs. Their ages were between 20 to 26. Although from the same university, our participants had experience (at least as interns) in more than a dozen of professional SE organizations, including Baidu, Microsoft, Tencent, Tiktok, etc. Therefore, the sample has reasonably good representativeness and the threats should be minimal. To ensure the randomness, they would draw a random number to determine their group assignment once they arrived at the experiment location, eventually formulating the overall group assignment in Fig. \ref{exp-de}.

\subsection{Experiment Process}

The experiment contained five steps. Upon the arrival of participants to the experiment room, they would be greeted by the research team first (Step 1). After deciding on the group assignment, one of the authors would brief participants on the study and the tasks they were going to finish (Step 2). Participants would have a few minutes to get themselves familiar with the environments (Step 3). Then they started working on the tasks with(out) ChatGPT's assistance until they felt comfortable to start (Step 4). After they started, they would have at most 75 minutes to finish the task. We made such a time restriction to keep the entire experiment within roughly one and a half hours. They could freely leave the experiment any time at their own will. They could also take breaks during the experiment but they were not allowed to communicate. Upon finishing the task, they would take a simple six-question survey (Step 5). The first five questions were used to measure the task load they experienced, and the last one was about their subjective perception of AI's effect in supporting software engineering tasks.

\begin{figure}[!h]
    \centering
\includegraphics[width=0.5\textwidth]{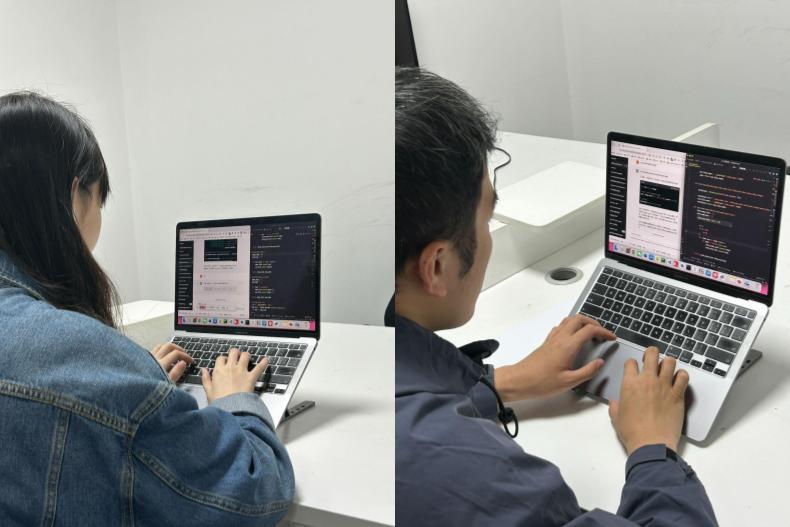}
    \caption{Two participants were in their experiment sessions, attempting to finish the tasks assigned to them.}
    \label{fig:my_label}
\end{figure}

Before leaving the experiment, all participants received a flat rate of 100 Chinese Yuan ($\approx$ 14.5 USD) as compensation. The compensation had nothing to do with their performances in the experiment. For those who performed best in the experiment, we gave them a small sum of bonus. Given that the entire experiment process was no longer than one and a half hours and most participants finished in about an hour, the effective hourly rate was actually well over 10 USD. This was 2.7 times of local minimal wage (25.3 Chinese Yuan) in Beijing, where UICTBC is located. It was also comparable to the junior software engineers' average salary. Therefore, the compensation satisfied the general ethical guidelines \cite{10.1145/3411764.3445734}. 

\subsection{Pilot Experiment}
To evaluate and improve the experiment design, we performed three pilot sessions before the formal execution of the experiment. The pilot sessions followed the same protocol as the initial experiment design except for not employing randomized assignment. During the pilot sessions, a few participants performed all the tasks. We observed their behaviors and performances, as well as collected their feedback. Then, we slightly adjusted the experiment design accordingly, e.g., lifting the compensation. The pilot experiment demonstrated that the design was generally effective.

\section{Data and Analysis}
\subsection{Collected Data and Measurements}
We collected both quantitative and qualitative data in the study. The quantitative data was metrics extracted from participants' sessions. Tab. 1 summarized the main variables and their meanings. There measurements were as follows. First, we measured the \emph{time} used to finish the experiment task. It reflected the \emph{Efficiency} of different task and workflow conditions. This metric was reverse-coded, since longer time used in fulfilling the task indicated less efficiency. Second, we measured the \emph{Solution Quality} of participants' work using a set of test cases. If all test case passed, the score would be 100, otherwise, the score was proportional to passed test cases. \emph{Subjective Perception} was directly extracted from the corresponding question in the post-study survey. \emph{Task Load} was measured by the widely-adopted \emph{NASA-TLX} \cite{hart2006nasa} in the post-study survey. 

\begin{table}[!h]
    \centering
    \caption{Main variables and their meanings.}
    \begin{tabular}{cl}
    \toprule
    \textbf{Variables} & \textbf{Meanings}\\
    \midrule
    \emph{Efficiency}     &  How fast a task could be solved by participants.\\
    \emph{Solution Quality}     & How good a participant's solution to a task is. \\
    \emph{Subjective Perception}  & How useful LLMs could be in development.\\
    \emph{Task Load} & How demanding a task is for a participant. \\
    \bottomrule
    \end{tabular}
   
    \label{tab:my_label}
\end{table}

The qualitative data was mainly drawn from the screen histories of the participants. These qualitative data offered us an opportunity to track fine-grain interactions between participants and ChatGPT, thus enabling us to answer the second research question. In addition, some participants' voluntary feedback in narratives was also a complementary qualitative data source. Since almost a half of participants did not work with ChatGPT, hence, their data was merely as references for comparisons with those with ChatGPT.

\subsection{Data Analysis Strategy}
Given the multi-type nature of the data, we combined quantitative and qualitative methods in analyzing the collected data. 

We used quantitative analysis to answer the $\mathbf{RQ_1}$. Recall that we had four experiment groups: (1) \emph{W-CP}, (2) \emph{W-TD}, (3) \emph{WO-CP}, and (4) \emph{WO-TD}. By comparing the variable's values of different groups, we could examine the benefits (or losses) of using ChatGPT. For example, suppose we want to know if using ChatGPT significantly increased efficiency in solving \emph{Coding Puzzles}, calculating the differences of \emph{time} used between group \emph{W-CP} and group \emph{WO-CP} would well serve the purpose. In practice, we found that most of those values did not follow normal distributions, even after data transformations, which violates the assumptions of commonly-used statistical methods such as ANOVA. Therefore, we used non-parametric method Mann Whitney U test (M-U Wilcoxon test) \cite{rey2011wilcoxon} to compare the differences. We calculated the effect size using the $Z$ statistics, as suggested by Vargha \& Delaney \cite{vargha2000critique}. All statistical analyses and visualizations were performed on a 2022 Macbook Pro with \textsc{R} statistical software (v 3.6.3).

Answering $\mathbf{RQ_2}$ required a detailed analysis of participants' screen history. Its qualitative nature determined that we should employ qualitative video data analysis methods. We followed the method used in Wang \cite{wang2017characterizing} to conduct detailed coding of interactions between participants and ChatGPT. For an interaction, it would be first coded in the following 4-tuple form: $<$\emph{start time}, \emph{end time}, \emph{participant's prompt}, \emph{ChatGPT's response}$>$. The context of each interaction is also coded and kept as memos. After coding all interactions, they were categorized to produce interaction categories. The categorization process was performed independently by two authors with an 0.92 IRR. All disagreements were resolved in a meeting of all authors. Then, different participants' activities were interconnected and compared to identify the patterns.  

\section{Results and Findings}

\subsection{$\mathbf{RQ_1}$: Benefits (or Losses) of Working with ChatGPT}
The first research question concerns the benefits brought by working with LLMs represented by ChatGPT. We focused on three aforementioned aspects, and reported on the results as follows.  
\subsubsection{Efficiency}
\begin{figure}[!h]
    \centering
    \includegraphics[width=0.5\textwidth]{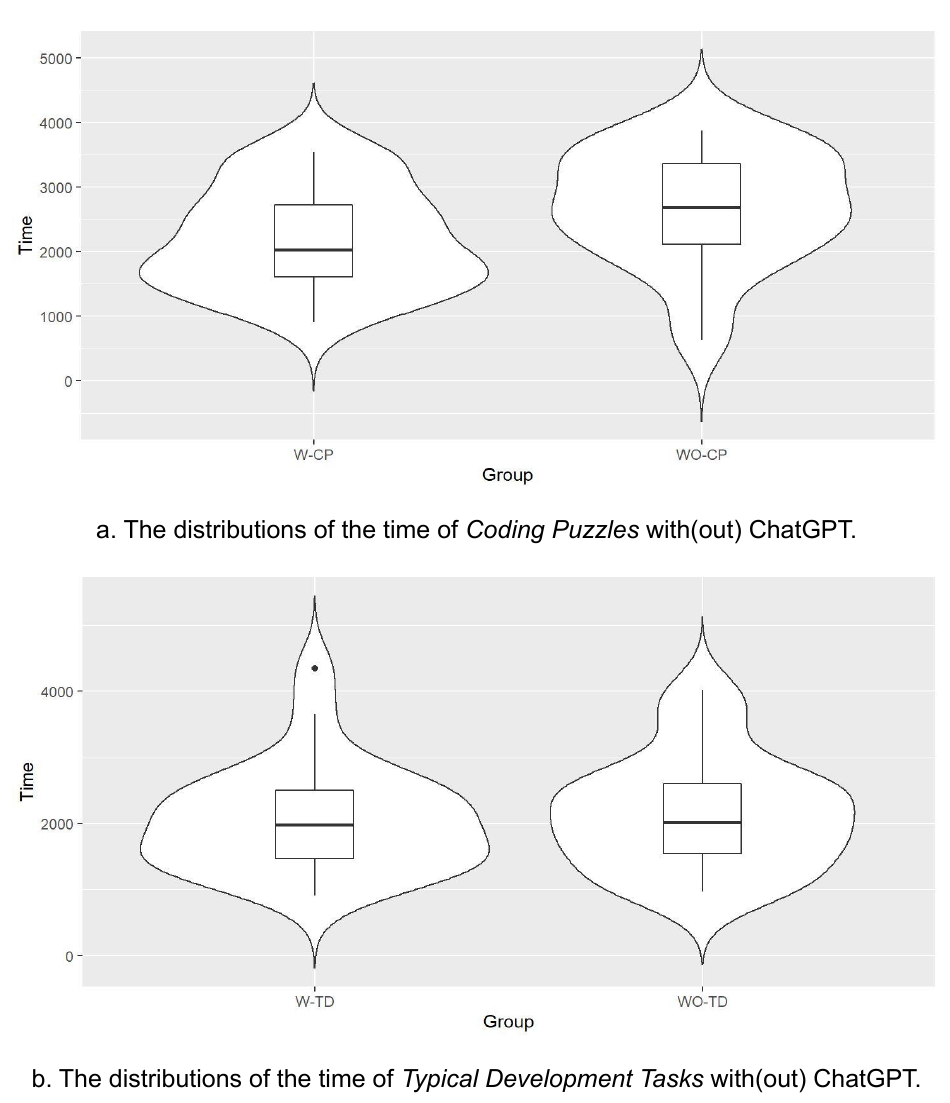}
    \caption{The distributions of participants' efficiency in different experiment groups.}
    \label{rq1-e}
\end{figure}

We first examine whether working with ChatGPT improved participants' efficiency in fulfilling the experiment tasks. Less time spent on the task means high efficiency, thus indicating high productivity. Fig. \ref{rq1-e} presented the distributions of participants' efficiency in fulfilling two tasks with(out) ChatGPT. First, let's have a look at Fig. \ref{rq1-e}.a; it was easy to find that those who worked with ChatGPT took less time to finish \emph{Coding Puzzles}. Those who worked with ChatGPT to fulfill the task had higher average efficiency (less time: 2196.7 vs. 2692.6 in seconds). The differences were also statistically significant, since the M-U Wilcoxon test results were \emph{W = 252} and \emph{p-value = 0.026}, with a moderate effect \emph{r = 0.30}. But that did not hold for \emph{Typical Development Task}. It was hard to find many visual distinctions in Fig. \ref{rq1-e}.b. While working with ChatGPT still yielded slightly less time: 2054.9 vs. 2147.3 in seconds, there was no statistical difference because the M-U Wilcoxon test results were \emph{W = 328.5} and \emph{p-value = 0.708}.  Therefore, we could draw a conclusion that:\textbf{Working with ChatGPT brought significant improvements in efficiency for Coding Puzzles, and slight yet insignificant improvements for Typical Software Development tasks}.

\subsubsection{Quality of Solutions}

\begin{figure}[!h]
    \centering    \includegraphics[width=0.5\textwidth]{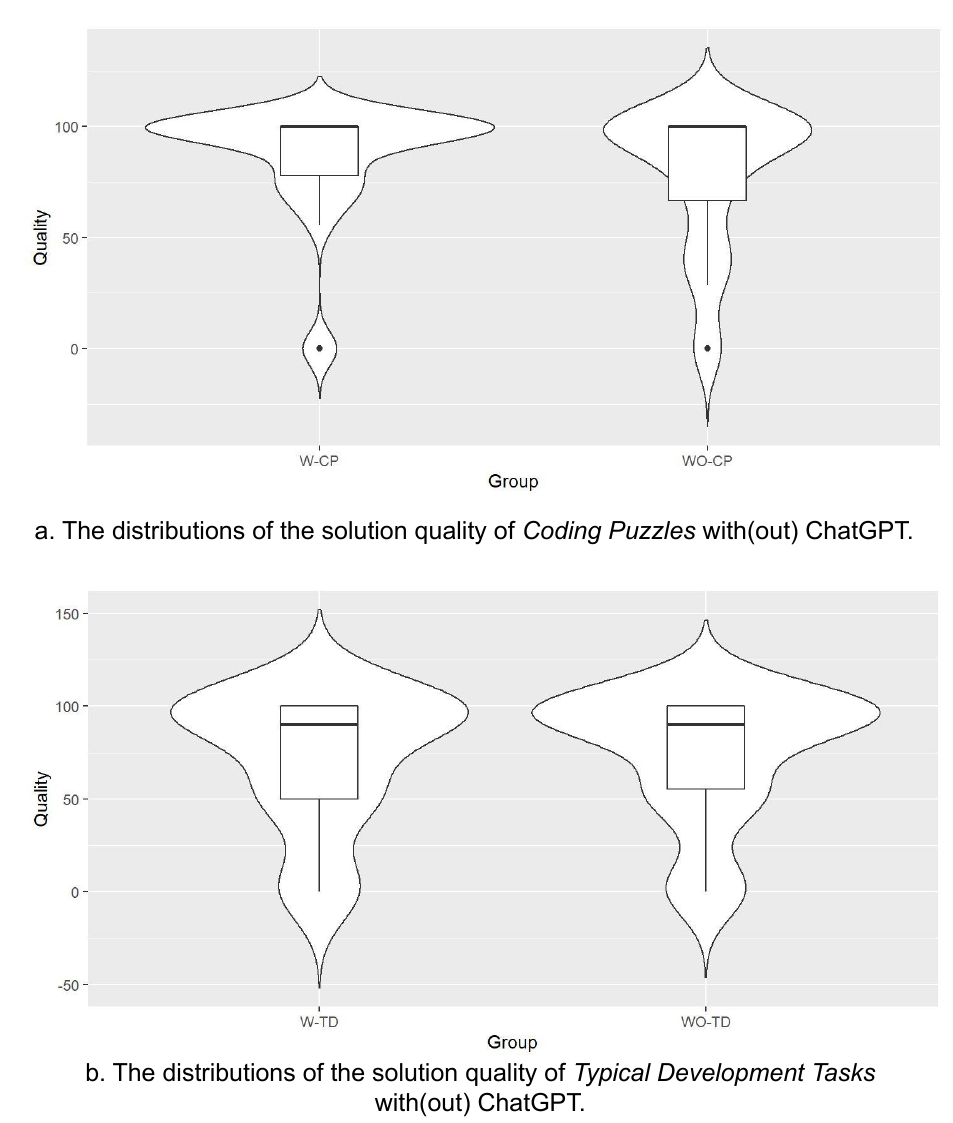}
    \caption{The distributions of participants' solution quality in different experiment groups.}
    \label{rq1-qu}
    \vspace{-1em}
\end{figure}
However, we found that using ChatGPT did not yield significant gains in the quality. Fig. \ref{rq1-qu} presented the quality of participants' solutions in two tasks with(out) ChatGPT. From Fig. \ref{rq1-qu}.a, we could find that the distributions were similar, except for the fact that those who work with ChatGPT obtained more consistent solution quality within the group. Those who worked with ChatGPT to fulfill the task had higher average quality (86.15 vs. 77.87). However, the differences were not statistically significant, since the M-U Wilcoxon test results were \emph{W = 434} and \emph{p-value = 0.39}. Also, Fig. \ref{rq1-qu}.b showed that the distributions of the quality of participants' solutions were almost identical for the Typical Development Task. However, for this task, those who did not work with ChatGPT had slightly lower average quality (73.68 vs. 71.46). Similarly, no statistical significance was detected with the M-U Wilcoxon test results \emph{W = 354} and \emph{p-value = 0.95}. Therefore, we could draw a conclusion that: \textbf{Working with ChatGPT might not yield better solution quality for both Coding Puzzles and Typical Software Development tasks}.

\subsubsection{Subjective Perception and Task Load}
Interestingly, most participants agreed that LLM tools such as ChatGPT would have a positive effect in helping them improve their productivity (see Tab. \ref{sub-per}), even though their direct experiences did not support this. I.e. in their subjective perceptions, ChatGPT should be a productivity booster. On the [0,20] scale, all participants' average perception is 16.12. Even the W-TD group (With ChatGPT in Typical Development Task) recorded a 14.36 average score, still falling into the medium positive spectrum. Therefore, it showed that: \textbf{People were generally positive to ChatGPT's effect, even when their direct experiences contradicted this}.

\begin{table}[!h]
    \centering
    \caption{Subjective perceptions of LLM tools' capabilities in improving productivity.}
    \begin{tabular}{ccc}
    \toprule
    \textbf{Groups} & \textbf{Mean} (Max. 20) & \textbf{SD}\\
    \midrule
      \emph{W-CP}   & 17.10  & 3.16\\
      \emph{W-TD}   & 14.36 & 4.86\\
      \emph{WO-CP}    & 17.20 & 2.66 \\
      \emph{WO-TD}   & 16.50 & 4.25\\
    \midrule
    All Participants & 16.12 & 3.97\\
    \bottomrule
    \end{tabular}  
    \label{sub-per}
    \vspace{-1em}
\end{table}

\begin{figure*}[!h]
    \centering
    \includegraphics[width=\textwidth]{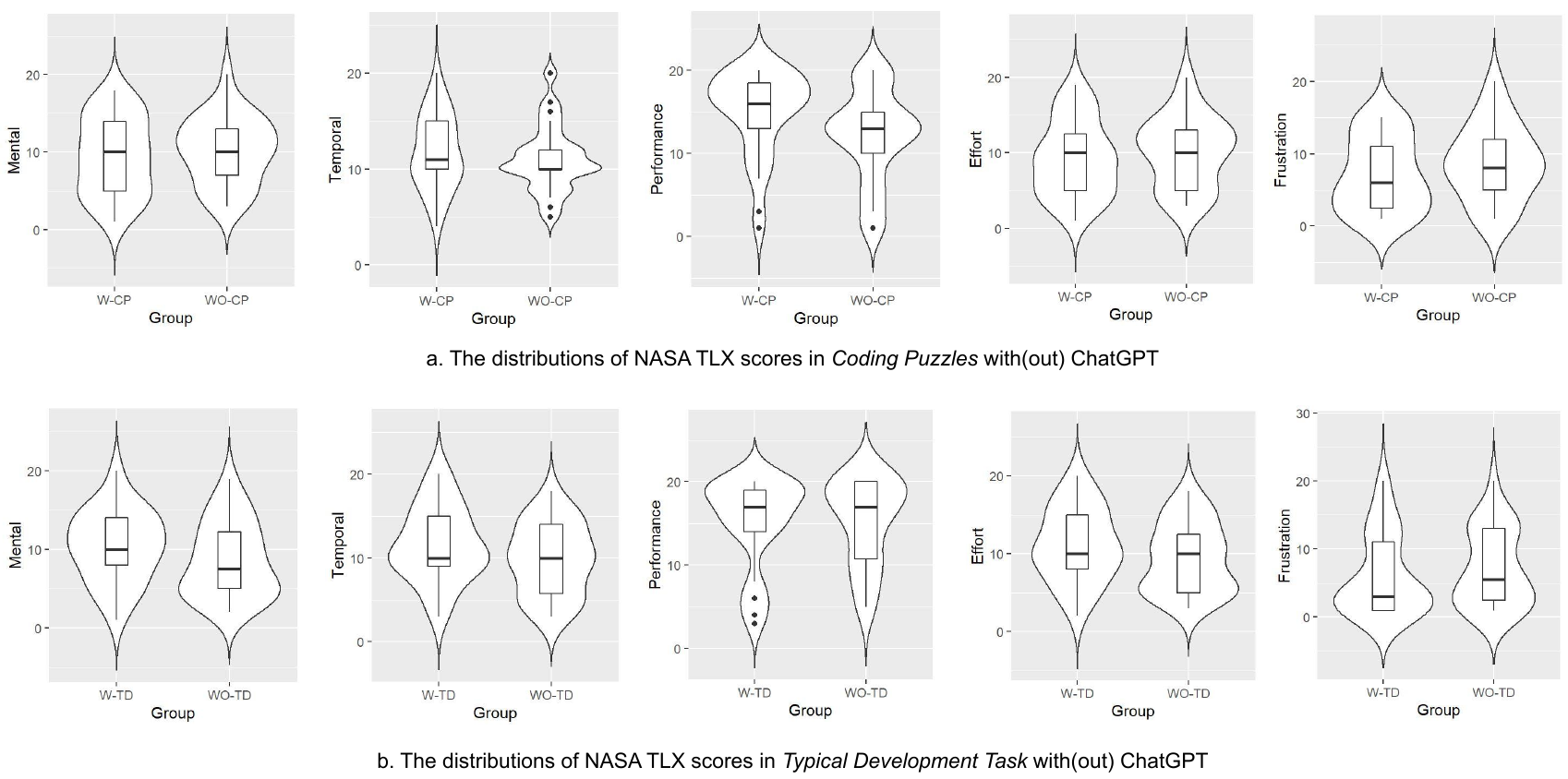}
    \caption{The distributions of participants' subjective task load in different experiment groups.}
    \label{nasatlx}
    \vspace{-1em}
\end{figure*}

Using the \emph{NASA TLX} data, we evaluated the effect of integrating ChatGPT into developers' workflow from five aspects of task load\footnote{We did not use the ``Physical Demand'' since it was not relevant to our experiment tasks which had minimal physical activities.}. Fig. \ref{nasatlx}.a described the distributions of \emph{NASA TLX} scores in \emph{Coding Puzzles}. Obviously, there were not many differences in three aspects--\emph{Mental Demand}, \emph{Effort}, and \emph{Frustration}. Their results from M-U Wilcoxon tests also did not suggest any significant differences. The \emph{Temporal Demand} and \emph{Performance} exhibited some visual differences. The former one did not obtain significant M-U Wilcoxon test results (\emph{W = 428}, \emph{p-value = 0.50}). However, the latter one's M-U Wilcoxon test results were significant (\emph{W = 524.5}, \emph{p-value = 0.023}) with a moderate to large effect (\emph{r = 0.50}). Fig. \ref{nasatlx}.b showed the distributions of \emph{NASA TLX} scores in \emph{Typical Development Task}. The distributions were similar for all five aspects, regardless of whether they were working with ChatGPT or not. Corresponding M-U Wilcoxon tests also support that. Therefore, we could conclude that: \textbf{Introducing ChatGPT to developers' workflows did not incur significant changes in their subjective task load for both experiment tasks, except it promoted people's perceptions of their performance when solving coding puzzles}. Furthermore, we think, based the conclusion from Ziegler et al. \cite{ziegler2022productivity}, ChatGPT's suggestions were accepted at a higher rate in coding puzzles, thus driving developer perceptions of productivity.

\subsection{$\mathbf{RQ_2}$: Interacting with ChatGPT}
The second research question concerns participants' interactions with ChatGPT in software development. Based on analyzing the screen histories, we found several significant patterns characterizing such interactions. We introduced them in detail as follows.
\subsubsection{Using ChatGPT in Coding Puzzles: Replacing Searching by Generating? Not for Everyone}

\begin{figure}[!h]
    \centering
    \includegraphics[width = 0.5\textwidth]{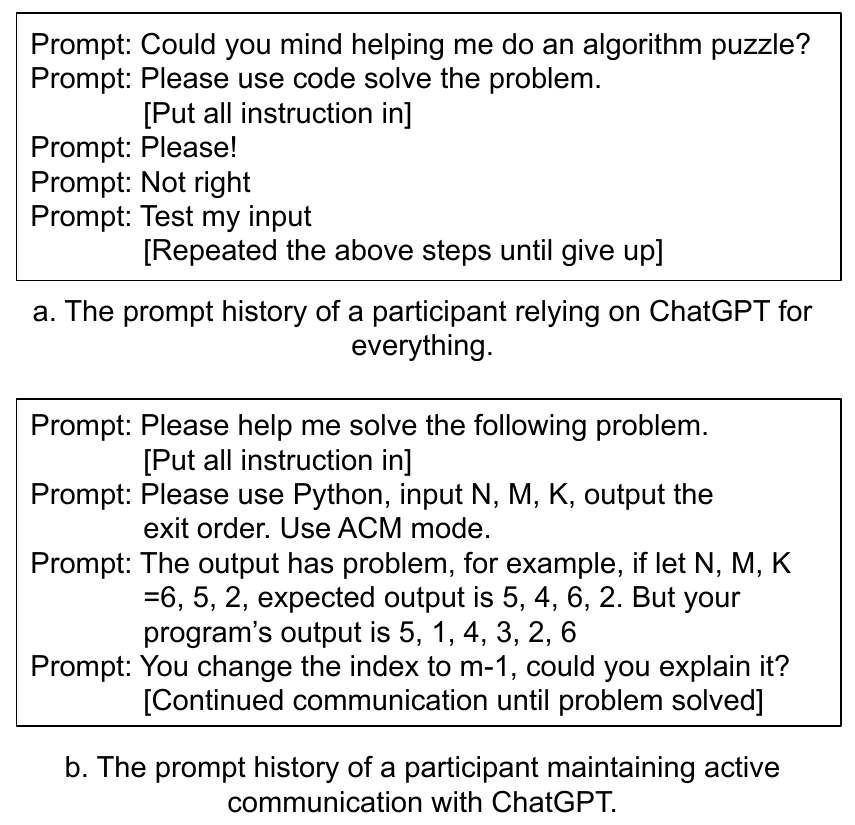}
    \caption{Examples of the prompt histories of the two different patterns.}
    \label{promptsexp}
  
\end{figure}

One of the typical work patterns of solving \emph{Coding Puzzles} was using it to replace code searching (8 out of 31). However, there were some divergences in their work processes. A few participants (5 out of 31) directly input the entire problem to ChatGPT without attempting to understand it. They relied on ChatGPT to do everything for them. They did not quite care what was generated. They would simply ``copy and paste'' the generated code. If there were any problems, they went back to ChatGPT to ask its help to sort them out. Fig. \ref{promptsexp}.a was such an example containing all the prompts the participant put in ChatGPT. A majority of participants were less passive (17 out of 31). While their interactions with ChatGPT started by letting it generate the initial code, they did not rely on it to make improvements. Instead, they fixed the code by themselves and asked specific questions to ChatGPT in the process. They spent a sinificant amount of session time reviewing and editing the code ChatGPT generated. Since ChatGPT responds with a delay, programmers delay thinking, and use the waiting time to think about GPT's historical reply. Fig. \ref{promptsexp}.b provided such a participant's prompts. This fits well with the interaction pattern proposed by Mozannar et al. \cite{mozannar2023reading}. Usually, they performed better than those who authorized ChatGPT to do all work for them.

There were still some participants (6 out of 31) tend to use online searching services to find the resources to help them solve the problem. They took an iterative cycle of \emph{coding}$\rightarrow$\emph{debugging}$\rightarrow$\emph{searching} to solve these puzzles. Those participants' work patterns exhibited few distinctions with these traditional problem-solving ways with Internet resources available reported in Wang \cite{wang2017characterizing}. Interested readers may refer to that paper for more details. 

Therefore, we could conclude that: \textbf{For tasks like Coding Puzzles, letting ChatGPT generate code first was replacing the traditional code searching practices, although there were still a few participants performing work in the traditional way}.

\subsubsection{Using ChatGPT in Typical Development Work: Treat It as My Life-Saver or A Colleague?}

We observed much richer and more complex interaction dynamics patterns when participants worked with ChatGPT. We summarized some key observations as follows. 

\begin{figure}[!h]
    \centering
    \includegraphics[width = \textwidth]{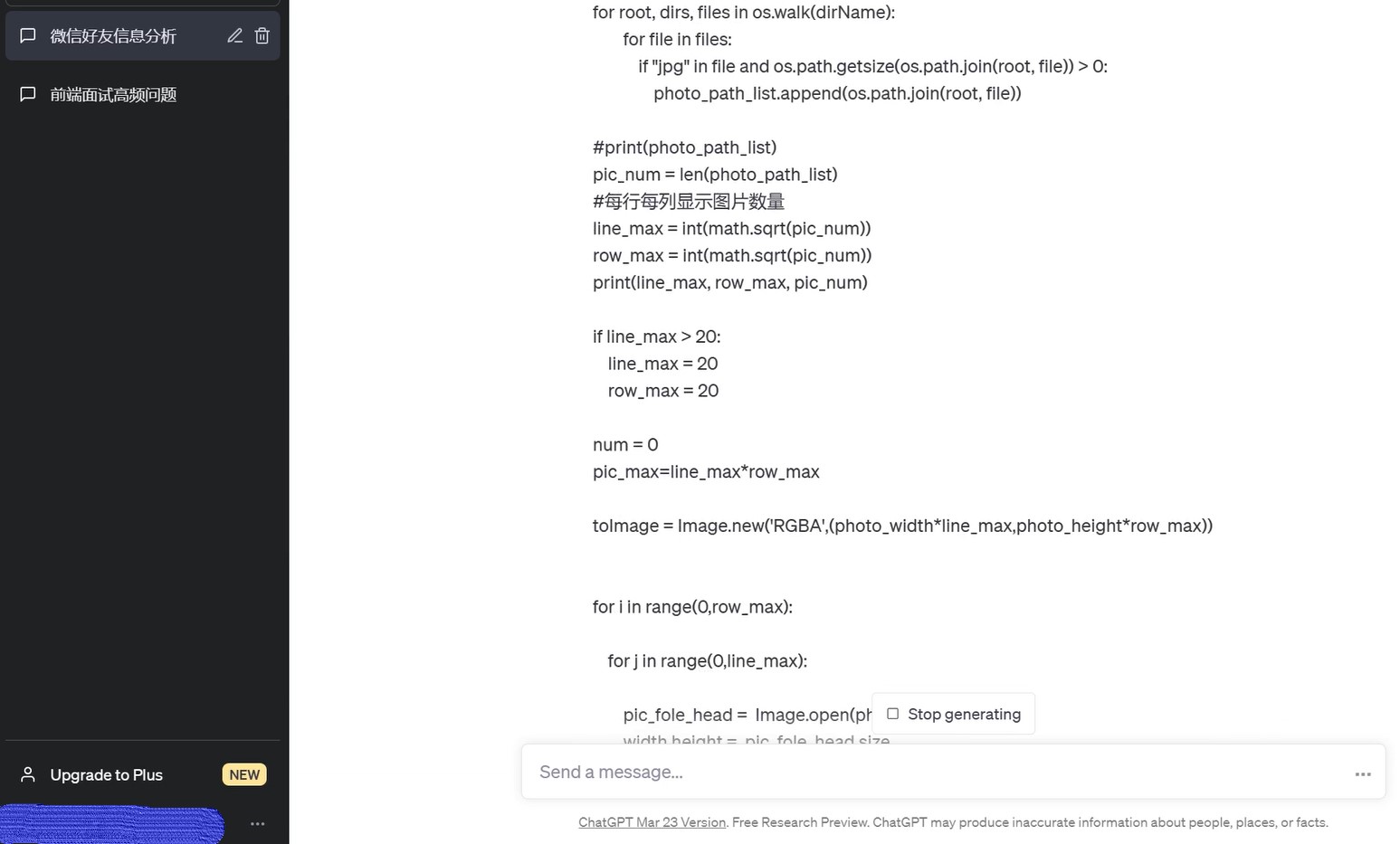}
    \caption{A participant put all the content of a buggy source code file into ChatGPT prompt as the input.}
    \label{allin}

\end{figure}

Although this task's instructions were much more complex, about one thirds (8 out of 25) of participants still attempted to let ChatGPT take charge of everything for them. They often asked a question in the form ``\emph{please help me fix it}'' or ``\emph{what should I do to make it run/right/...}'', and threw all instructions and code in a file into ChatGPT together with their requests. Fig. \ref{allin} provided such an example. ChatGPT usually gives some feedback such as ``\emph{Doing [...] may help you [...]}'' and presents some instances. Obviously, these were not what the participants wanted. They then tried some tricks, e.g., breaking code into several parts, or asking ChatGPT to locate the bugs, etc. However, they would never try to understand the task by themselves. Eventually, they often had to give up or ended with a very low-quality solution.

\begin{figure}[!h]
    \centering
    \includegraphics[width=0.6\textwidth]{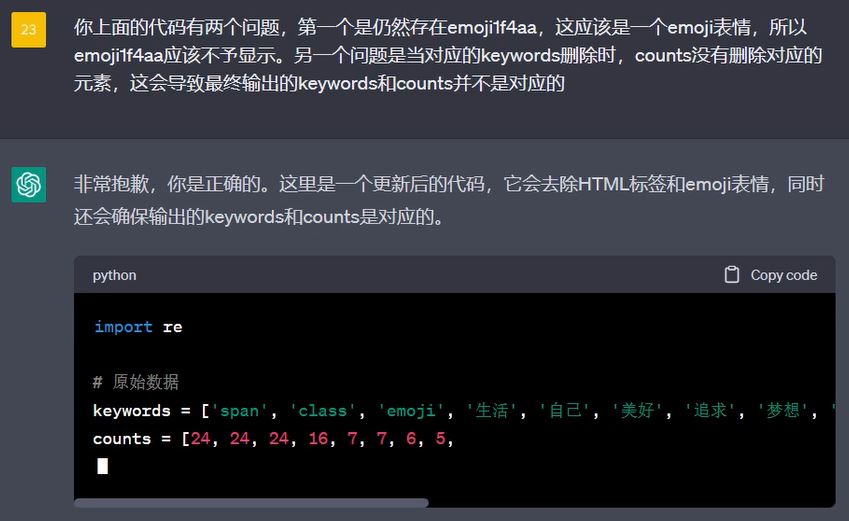}
    \caption{A participant discussed with ChatGPT about issues in a potential solution (partially in Chinese, see the above paragraph for explanations).}
    \label{discuss}
\end{figure}

Almost 50\% participants took another strategy (12 out of 25). They interacted with ChatGPT to get help throughout the entire process of their work but for different purposes. They began by trying to understand the source code. However, they asked ChatGPT's help to understand the code by letting ChatGPT figure out some code snippets' functions. After understanding the code, they attempted to develop a solution. In this process, ChatGPT was treated as a \emph{colleague}. They exchanged ideas and discussed with each other. These participants were those who best explored and utilized ChatGPT's cognitive and conversational nature. Fig. \ref{discuss} presented such an example. In this example, ChatGPT suggested a solution. However, the participant noticed there was an issue and told ChatGPT, ``\emph{Your code has two issues. First, the emoj1f4aa still exists. It is an emoji, so it should not display. Second, when corresponding keywords are deleted, elements in counts do not delete accordingly. This led to inconsistencies between the keywords and counts in the final output}[translated from Chinese]. ChatGPT agreed and said ``Very sorry, your are right[...]'' It also prepared a new version where the two issues were solved for the participant to consider. 

A few participants (5 out of 25), often those who had good software development capabilities, also started their workflow by understanding the source code first. Instead of interacting with ChatGPT to understand the source code, they mostly did the program comprehension by themselves. Then, they asked ChatGPT some very specific questions, for example, programming libraries which usually were some questions that could be addressed by searching. They often completed the task at a very fast pace. In the process, they kept minimal use of ChatGPT.

Therefore, we could conclude that: \textbf{Participants' interaction with ChatGPT exhibited rich dynamics. Some of them viewed it as a colleague and developed complex interactions to get helps from ChatGPT, while some other struggled to work with it.}

\subsection{Summary of Findings}
Based on the above results, the main findings could be summarized as follows:
\begin{itemize}
    \item Using ChatGPT brought significant improvements in efficiencies for Coding Puzzles but not for Typical Development Tasks. [$\mathbf{RQ_1}$]
    \item Working with ChatGPT might not yield better solution quality for both Coding Puzzles and Typical Software Development tasks. [$\mathbf{RQ_1}$]
    \item Introducing ChatGPT into development workflows did not incur an extra task load, and participants generally held positive attitudes about such tools. [$\mathbf{RQ_1}$]
    \item ChatGPT might have the potential to replace the code searching practices for simple Coding Puzzles. [$\mathbf{RQ_2}$]
    \item Participants' interaction with ChatGPT exhibited rich dynamics. Some participants were quite struggling, while almost half of them developed highly-interactive strategies to work with it in dealing with Typical Development Tasks. [$\mathbf{RQ_2}$]
\end{itemize}

\section{Discussions and Implications}

\subsection{Discussion of the Findings}

In his classic ``No Silver Bullet''  paper \cite{brooks1987essence}, Fred Brooks classified the difficulties in software engineering into two categories, essence, and accidents, and wrote down:
\begin{quotation}
  \noindent \emph{I believe the hard part of building software to be the specification, design, and testing of this conceptual construct, not the labor of representing it and testing the fidelity of the representation. We still make syntax errors, to be sure; but they are fuzz compared to the conceptual errors in most systems.}
\end{quotation}
Our findings, to some degree, reconfirmed Brooks' insightful arguments. Yes, LLMs such as GPT could generate grammatically-perfect code, while our humans still made syntax errors. As $\mathbf{RQ_1}$'s findings demonstrated, it did significantly improve the efficiency in solving \emph{Coding Puzzles}, confirmed prior findings in literatures such as Kazemitabaar et al. \cite{10.1145/3544548.3580919}. However, when dealing with tasks simulating real-world development, LLM's capabilities became limited since such tasks usually required to establish conceptual models of the software system.
As $\mathbf{RQ_2}$'s findings demonstrated, participants who trusted the AI less and were more involved in the format of the prompt got more predictable responses, which supports the conclusion of Perry et al. \cite{perry2022users} and Sandoval et al. \cite{sandoval2023lost}. 
$\mathbf{RQ_2}$'s findings documented fine-grained interactions of Human-AI collaboration in SE domain. They echoed recent literature such Weisz et al. \cite{10.1145/3490099.3511157}
by emphasizing that human-AI collaboration, rather than overly relying on one party for better outcomes, but in a greater detail. Besides, we also identified LLMs' possible potentials in replacing source code searching tools. 

\subsection{Implication to Research}
Along with the findings, our study also suggested rich future research opportunities. First, we identified some SE areas in which the current LLMs needed to be more sophisticated to provide meaningful support to SE activities. Researchers might need to conduct more empirical studies to reveal more and to develop SE-specific. Second, we found rich dynamics when participants collaborated with LLMs tools. We believe that much more dynamics were still uncovered since we only focused on a very limited number of tasks, particularly the tasks requiring multimodal interactions \cite{akata2020research}, for example, generating architectural design diagrams. Researchers, including ourselves, should continue this journey of discoveries. Moreover, CSCW researchers had done a bunch of work in human-AI collaboration, including theories, empirical studies, tools, etc., and were developing methodologies tailored for future investigations, e.g., \cite{10.1145/3555212,wang2019human, 10.1145/3334480.3381069}. SE researchers should team up with them to jointly advance the progress in this direction. Lastly, the integration of LLMs into the development workflow was non-trivial. Even for the small tasks in our study, we noticed someone benefited from it while some others were trapped in it (See \S 5.2). How to design interaction mechanisms, personal \& team workflow templates or guidelines, and human-AI collaboration tools, to make as many as possible software engineers benefit from it should be an important future topic. 

\subsection{Implication to Practices}

There are basically two alternative future worlds. In one world, large language models and other AIGC techniques would replace human developers and end conventional software development. In such a world, most software will eventually be replaced by AI models which could directly execute that task after being given an appropriate description of a task. The design, development, or maintenance of conventional software would no longer exist. Educating software engineers thus becomes unnecessary. In another future world, these models may not evolve fast enough or never develop sufficient cognitive and task execution capabilities to eliminate human involvement. Human effort will still be necessary in order to build real, usable systems. 

In such a world, working with AI shall be an essential part of professional software engineers' skillset. Part of software engineering activities might be so-called ``prompts engineering''\cite{white2023prompt}, which has dual meanings here. First, software engineers might need to learn and excel in how to articulate good prompts to let LLMs do what they mean to do. However, we had very limited knowledge and cues about how such LLMs work and evolve internally. A possible way was to encourage a collective ``\emph{trial and error}'' practice, where individual developers report their successes and failures in collaborating with LLMs, then compile tutorials from the collected experiences. Second, it also requires software engineers to think and design effective prompts systems, which have yet to receive limited attention in the SE community.

LLMs might offer some potential solutions to the long-standing diversity problem in the software engineering profession. They reduced the amount of specialized knowledge required to develop complex software systems; that would enable a broader range of individuals to become involved in software engineering.

\subsection{Implication to SE Education}
The penetration of LLMs into software development and its future evolution would have profound implications for how we, as educators, train future-generation software engineers. There were many open questions regarding future SE education remaining unanswered, or we were not even clear if these questions were valid. To name a few, What should be the core concepts and knowledge of software engineering? Shall we transform the current programming-based curriculum to some other things? How could we reinforce the links between design from multiple views and source code in a possible world where most code is generated by AI? What will future software development look like? How to properly assign responsibilities between human developers and AI developers? ... Thus, the SE education community might need to develop an agenda to systematically evaluate LLMs' impacts on the SE curriculum, identify the challenges, and address them with the entire community's collective efforts.  

\section{Threats to Validity}

From the perspective of \emph{construct validity}, most constructs used in this study were well-defined and measured by objective metrics obtained from direct observations. For example, the efficiency was directly measured by the time spent on the given task. Only one contract--\emph{Task Load}--was not directly measured. Although we did not use some bio-meters to directly measure it, we adopted a well-established measurement system (\emph{NASA-TLX}) as the indicators and strictly followed its user guidance. \emph{NASA-TLX} had been used by thousands of studies, including those similar to our settings \cite{10.1145/2568225.2568266,wang2017characterizing}. Therefore, we are confident that there was a minimal level of threats to construct validity.   

From the perspective of \emph{internal validity}, similar to any human-subject laboratory experiment, we must admit that we could not perfectly control all relevant factors. For example, there might be differences among experiment groups in their software development abilities. However, standard randomization assignment eliminated plenty of such concerns. In addition, all experimental groups were assembled initially and maintained throughout the study. Besides, the same normalized measurement instruments were used and applied to all groups in an unbiased fashion. The main treatments were straightforward. All finished tasks with the same language. The data analysis and reporting followed standard guidelines. Thus, the execution of the study achieved good internal validity \cite{rugge2012screening}. However, many other factors, such as unintentional plagiarism, personal differences attitudes, working environments such as setting dual-display, etc., may still play a role in the study. 

From the perspective of \emph{external validity}, we must admit that our study inherited most threats to the external validity of any controlled experiments \cite{feitelson2022considerations}. For example, we only studied one type of task, rather than hundreds of kinds of tasks in the entire lifecycle of software development. Besides, our study is also relevant to the long-standing debate about using students in software engineering experiments \cite{falessi2018empirical,10.5555/2818754.2818836}. However, we required participants to have at least one year experience in professional development organizations and most participants (102 out of 109) were graduate students. Therefore, we argue that their behaviors could partially simulate professional developers'. Meanwhile, the main purposes of the study is to examine ChatGPT's capabilities. Thus, students' capabilities might not be a threat \cite{10.5555/2818754.2818836}. Indeed, As an elite university in ICT, the technical skills of UICTBC's students  were widely assumed to be above the average software engineer. Moreover, the AIGC tools for SE tasks keep evolving, future generations of such tools may be much more powerful than the GPT 3.5 series of models used in this study. Anyway, researchers shall be cautious about extending the findings of this study to other settings. 

\section{Concluding Remarks}
It has yet to be clear if LLMs or other AI technologies will eventually replace software development professionals. However, working with these models to produce software seems inevitable in the foreseeable future. From a human-centered perspective, this paper reports on a controlled experiment evaluating the capability of LLMs represented by ChatGPT with 109 participants. The experiment revealed that the state-of-the-art LLMs still needed to develop sufficient capabilities to fully support Typical Software development activities while they did help improve the efficiency in solving Coding Puzzles. Using the screen history data of participants' interactions with ChatGPT, this study also characterized rich dynamics and patterns of how people collaborated with AI. We found that continuous, purposeful, and specific communication with ChatGPT might resulted some positive effects in performing software engineering activities.

As the first large-scale experimental evaluation of LLMs' capabilities in software development, this study \emph{rebutted} the claim that LLMs would soon replace software developers in typical software development tasks. Even for very simple tasks in typical development scenarios, such as fixing simple bugs in our experiment, ChatGPT did not exhibit any sign of overwhelming superiority, i.e., it neither improved efficiency nor brought better quality. However, there were signs that sophisticated use and articulate input might be beneficial. In a foreseeable future, we could expect more collaboration between developers and such LLMs. As researchers, we must further build theoretical and empirical knowledge on such collaboration, and design mechanisms and tools to promote effective collaboration. Moreover, it is also a good timing for us to rethink how future software engineers should be trained to better work with future AIs.  

\begin{acks}
The authors would like to thank all participants of the study, as well as the anonymous reviewers for their insightful comments and suggestions. This work is partially supported by National Natural Science Foundation of China (NSFC) under grants 62076232 and 62172049. Corresponding author: Yi Wang.
\end{acks}

\section*{Data Availability}
The experiment protocol and materials are permanently available at figshare: \url{https://doi.org/10.6084/m9.figshare.24166248}. The source code for Task 2 (typical development task) is included. The anonymous dataset used in the quantitative data analysis is also enclosed. However, due to privacy considerations, the raw materials, e.g., videos and screen histories, will be available upon requests from researchers who have already been granted IRB approvals from their own institutions.

\appendix
\section{The Five Coding Puzzles Used in the Experiment}
The five coding puzzles used in Task 1 (\S3.1.1) are listed as follows. Note that a participant of Task 1 only need to solve two puzzles randomly drawn from them.
\subsection{Project Leader-board}
Suppose each student's information includes ``\emph{Username}'', ``\emph{Total Progress}'', and ``\emph{Total Number of Problems Solved}.''. In the progress leader-board, generate a leader-board according to \emph{Total Progress} and \emph{Total Number of Problems Solved}. First, input the information of $n$ students; then sort them in descending order according to \emph{Total Progress}. For students make the same \emph{Total Progress}, sort them in descending order according to their \emph{Total Number of Problems Solved}. When students have the same \emph{Total Progress} and \emph{Total Number of Problems Solved}, their rankings would be the same. However, when outputting the information, they would be listed in ascending order by ``\emph{Username}''.

\noindent \textbf{Input Format}:
First, enter an integer $T$, which means the number of test data sets, followed by $T$ sets of test data. Each set of test data first inputs a positive integer $n$ ($1 < n < 50$), which represents the total number of students. Then, enter $n$ lines, each line includes a string $s$ (no more than 8 characters) without spaces and two positive integers $d$ and $t$, which represent the username, total progress, and total number of problems solved respectively.

\noindent \textbf{Output Format}:
For each group of tests, output the final ranking. Each line contains one student's information, including rank, username, total progress, and total number of problems solved. Leave a space between each piece of data on each line. If both the total progress and the total number of problems solved are the same, then the ranking is also the same; otherwise, the ranking is the corresponding sequence number after sorting.

\subsection{Analyzing Valve Status in the Workshop}
The CPU reads one byte of content through an 8-bit IO port, which is now stored in a bytes object, for example: b'$\backslash$x45'. These eight bits represent the current status of eight valves in the workshop, where 1 indicates that the valve is open, and 0 indicates that the valve is closed. Please write a program to parse the current status of the eight valves in the bytes object, where ``\emph{True}'' means open and ``\emph{False}'' means closed. These eight statuses should be organized in a list, where the $i$-th element corresponds to the $i$-th bit of the input byte. Write a program to implement the above analysis.

\noindent \textbf{Input Format}:
A single byte in the form of b'$\backslash$x45'.

\noindent \textbf{Output Format}:
A list containing eight Boolean values, for example: [\emph{True}, \emph{False}, \emph{True}, \emph{False}, \emph{False}, \emph{False}, \emph{True}, \emph{False}].

\subsection{EAN13 Barcode Verification}
The packaging of the products purchased in supermarkets will have a two-dimensional barcode, generally an EAN13 code consisting of 13 digits. The first 12 digits include the country code, manufacturer code, and product code, and the last digit is a check digit, structured as follows ($N1$ to $N12$ are the first 12 digits, $C$ is the check digit):
``$N1 N2 N3 N4 N5 N6 N7 N8 N9 N10 N11 N12  C$''
The relationship between the check digit C and the first 12 digits is as follows:
Let:
$C1$ be the sum of the odd-numbered digits in the first 12 digits, i.e., $C1= N1+N3+N5+N7+N9+N11$. $C2$ is the result of multiplying the sum of the even-numbered digits in the first 12 digits by 3, i.e., $C2 = (N2+N4+N6+N8+N10+N12)\times3$.
$CC$ is the units digit of ($C1 + C2$). Then, $C$ is the units digit of 10-CC.
Write a program to check whether a 13-digit number entered by the user can pass the EAN13 code verification. 

\noindent \textbf{Input Format}:
Enter a 13-digit barcode.

\noindent \textbf{Output Format}:
If the barcode entered is not 13 digits, output ``Barcode error.'' If the barcode entered is 13 digits and can pass the verification, output ``Verification passed.'' If it is 13 digits but cannot pass the verification, output ``Verification failed.''

\subsection{Survival Test on a Pirate Ship}
A fully loaded ship is sailing at sea when it is suddenly surrounded by pirate ships. The pirates hijack the ship and pull all the crew ($N$ people in total) onto the deck. The pirate captain says: ``\emph{We will play a game, and the winners can leave. The rules are: $N$ people form a circle, start counting from the first one, the $M$-th person will be out of the circle; then continue the counting process, kick out every $M$-th person until there are $K$ people left.}'' Write a program to determine the order of leveling the circle and who will be left for given $N$, $M$, and $K$.

\noindent \textbf{Input Format}:
The values of $N$, $M$, and $K$ are given in one line, separated by spaces. The input guarantees that the data are all positive integers, and that $N > K$.

\noindent \textbf{Output Format}:
Output the order of safely leaving the circle. For example: if $N = 6$, $M = 5$, $K = 2$, then the order of leaving the circle is 5, 4, 6, 2, while 3, and 1 are left.

\subsection{The Simplest Word}
There are a large number of words, and learners grade these words based on their difficulty. A certain number of learners grade the 10 words they are learning, and different people may grade the same word. If a word is graded by multiple people, its composite score is the average of all grades. Write a program to find the word with the lowest composite score.

\noindent \textbf{Input Format}:
The first line $T$, indicating there are $T$ sets of data.
For each set of data, the first line $N$, indicating there are $N$ people. Then, input $N\times10$ lines, each line has a word and an integer (representing the score), separated by a space.

\noindent \textbf{Output Format}:
For each set of data, output the simplest word.

\bibliographystyle{ACM-Reference-Format}
\bibliography{sample-base}

\end{document}